\documentclass{ws-rv9x6}
\usepackage{ws-rv-van}             % numbered citation/references (default)
\makeindex
\def\tr{{\mbox{tr~}}}

\def \Jcrit{{\left(zJ/U\right)_{\textrm{c}}}}

\def \bk{{\bf k}}

\def \av#1{{\langle#1\rangle}} \def \bra#1{{\langle #1|}} 

\def \abs#1{{\left|#1\right|}}

\newcommand{\ket}[1]{\ensuremath{\vert #1 \rangle}}%

\newcommand*{\bfrac}[2]{\genfrac{}{}{0pt}{1}{#1}{#2}}

\def \r{ {\bf{r}} } \def \k{ {\bf{k}} } \def \R{ {\bf{R}} }

\begin{document}

\chapter{Quantum gases in optical lattices\label{chapterKollath}}

\author{Peter Barmettler}
\address{D\'epartement de Physique Th\'eorique,
 Universit\'e de Gen\`eve, CH-1211 Gen\`eve, Switzerland.}
\author[P. Barmettler and C. Kollath]{Corinna Kollath }
\address{HISKP, Universit\"at Bonn, Nussallee 14-16, D-53115 Bonn,
Germany}

\begin{abstract}
The experimental realization of correlated quantum phases with ultracold gases
in optical lattices and their theoretical understanding has witnessed
remarkable progress during the last decade. In this review we introduce basic
concepts and tools to describe the many-body physics of quantum gases in
optical lattices. This includes the derivation of effective
lattice Hamiltonians from first principles and an overview of the emerging
quantum phases. Additionally, state-of-the-art numerical tools to
quantitatively treat bosons or fermions on different lattices are introduced.
\end{abstract}

\body
\section{Introduction}
With the realization of Bose-Einstein condensates (BEC)
\cite{AndersonCornell1995,BradleyHulet1995,Davis1995} and quantum degenerate
Fermi gases \cite{demarco-1999,Schreck2001, Truscott2001} a new field of
research opened up, in which quantum phenomena driven by the interplay of a
macroscopic number of atoms are the focus. Early experiments explored mainly
coherent effects in dilute and weakly interacting Bose-Einstein condensates.
More recently,  with the realization of effectively strong and tunable
interaction between atoms, correlated many-body phenomena gained in interest.
The increase of atom-atom interaction has been achieved following two different
routes: On the one hand, Feshbach resonances  have been employed to directly
tune the effective scattering length of the atoms \cite{Bloch2008}.  On the
other hand, atoms have been loaded into periodic lattice potentials in which the
suppression of the tunneling leads to an effective enhancement of the
interaction \cite{Jaksch1998,Bloch2008}. This effective enhancement was
reported in the observation of number squeezing in a Bose-Einstein condensate
\cite{Orzel2001}. Subsequently, various confining laser geometries have been
developed to design different interesting situations.

Theoretically, atoms in optical lattices are often well described by so-called
(Bose)-Hubbard type models \cite{Jaksch1998}. In these lattice models the
intriguing interplay of quantum kinetic processes and local interaction can be
investigated in its cleanest form. Hubbard-type models have been studied over
decades in the context of solid state physics. However, for solids they are
often only a rough approximation which covers at best the most important
physical effects. Typical deviations are for example due to the long range
nature of the Coulomb interaction, the presence of defects, or lattice
vibrations. In contrast, cold atomic gases are very clean realizations of these
models and internal parameters, such as the geometry and interaction strength, are
tunable to a large extent.  The effective parameters can be derived from
microscopic principles.  Thus, cold gases in optical lattices offer the
possibility to simulate the physics and to test theoretical predictions of
Hubbard models in a controlled way. For example Bloch-oscillations, which were
predicted for electrons in solids, but difficult to access due to disorder
effects, were observed in cold gases \cite{Peik1997a}. In another remarkable
precision experiment the evolution of the Fermi surface with increasing density
was imaged in non-interacting gases \cite{KoehlEsslinger2005}.  With bosons in
cubic optical lattices \cite{Greiner2002a}, effective interactions were
increased such that a phase transition between a superfluid to a
Mott-insulating state was observed. This interaction driven phase transition is
nowadays one of the most thoroughly investigated examples of quantum emulation
of an interacting many-body system. Experimental setups for studying more
complex interaction driven phenomena such as charge density waves,
superfluids, and topological phases are subject of current investigations.
Important challenges for these ambitious aims are efficient cooling schemes and
the realization of non-cubic geometries such as superlattices, frustrated
geometries or honeycomb lattices.

In this brief review we would like to introduce basic theoretical concepts and
tools for the description of many-body phases which can be accessed using cold
atoms. We mainly focus on s-wave interactions between atoms and deep optical
lattice potentials. In such situations most of the physics is contained in the
Hubbard-like models. We derive this model in Sec.~\ref{sec:KollathBasic} and
discuss the superfluid to Mott insulating transition in a basic mean field
formalism in Sec.~\ref{sec:KollathPhases}. In the third part
\ref{sec:KollathMethods} we describe powerful numerical tools which can be
adopted to describe interacting cold atoms.  Some advanced topics and
extensions to the standard Hubbard models are discussed in Sec.
\ref{sec:KollathOutlook}.

\section{Basic description of ultracold gases trapped in optical lattices}
\label{sec:KollathBasic}
\subsection{Optical lattice potentials}
Off-resonant laser light can be used to form effective conservative potentials
for neutral atoms due to the so-called Stark effect, also called light shift (See Ref. 
\cite{FootBook} or chapter 10). This effect is based on the induced dipole interaction between
neutral atoms and electromagnetic waves. Applying an off-resonant electric field ${\bf
E}(\r,t)$, the effective potential strength felt by the atoms is given by
$V(\r)=-\frac{1}{2}\alpha |{\bf E}(\r)|^2$. Here $|{\bf E}(\r)|^2$ is the
intensity of the light field averaged over a time much longer than the period
of the electromagnetic wave. The polarizability $\alpha$ of the atoms 
depends on the induced electric dipole moment and is proportional to
the inverse detuning $-1/\delta$ of the frequency of the light field from the atomic
transitions\footnote{There are also non-conservative contributions to the
optical lattice potential, which, however, become negligible sufficiently far
from resonance.}. Thus, the sign of the potential can be changed from
attractive to repulsive by changing the detuning from red to blue detuned.

The dependence of the effective potential on the intensity of the light field
enables the realization of a great variety of distinct geometries. Using for
example a simple retroreflected laser beam with wavelength $\lambda$ (and wave
vector $k_L=2\pi/\lambda$) along the direction $x$, a standing wave can be
generated. This leads to a periodic potential of the form $V(x)=V_{x}
\sin^2(k_Lx)$. The lattice period $a$ is given by half of the wavelength
$\lambda/2$ and in typical experimental setups $a$ lies between $100 nm$ and
$5\mu m$. It is convenient to measure $V_x$ in units of the recoil energy
$E_R={\hbar^2k_L^2}/{2m}$, where $m$ is the atomic mass. The recoil energy is
typically of the order of a few kHz.  In addition to the periodic structure,
the focus of the laser beam gives an overall potential in particular in the
orthogonal direction to the beam propagation which we have neglected. 

More complicated lattice structures can be generated by the combined
application of several laser beams, i.e. ${\bf E}(\r,t)=\sum_i {\bf
E}_i(\r,t)$. Often used are cubic lattices. These are formed by three
orthogonally polarized  standing laser beams along orthogonal spatial
directions. The resulting potential can be described by
$V_L(\r)=\sum_{\nu=x,y,z} V_{\nu} \sin^2(k_L r_\nu)$, where $r_\nu=\r\cdot{\bf
e}_\nu$ is the projection onto the primitive vectors of the cubic lattice.
This rather simple setup offers the possibility to change the geometry between
a three dimensional crystal and arrays of planes or tubes by tuning the
intensities of the laser beams in the different directions.  Various extensions
such as more complex geometries, state-dependent potentials, or superlattices
have been realized already.
A separate chapter of this volume is devoted to the experimental realization of
special optical lattices (chapter 5).

\subsection{A single particle in a periodic potential} 
In order to understand the rich physics induced by optical lattice structures,
it is useful to revise the motion of a single particle in a perfectly periodic
potential\cite{Bloch1929}. Interactions and inhomogeneous contributions will be
taken into account subsequently. We consider a potential $V_L(\r) =V_L(\r+\R)$,
where $\R$ is a lattice vector. According to Bloch's theorem
\cite{Bloch1929,Ashcroft1956}, the single-particle eigenfunctions, the
so-called Bloch functions, can be represented by $\phi_{\k}^{(n)}(\r)= e^{i\r
\k} u_\k^{(n)}(\r) $, a product of a plane wave and a function
$u_\k^{(n)}(\r)$. The function $u$ has the same periodicity
$u_\k^{(n)}(\r)=u_\k^{(n)}(\r+ \R)$ as the lattice potential. Index
$n$ labels different Bloch bands. The quasi-momenta $\k$ are related to
eigenvalues of translations by lattice vectors $\R$ and lie within the first
Brillouin zone (BZ), i.e.~$(-\frac\pi a,-\frac\pi a,-\frac \pi a)\leq \k <
(\frac \pi a,\frac \pi a,\frac \pi a)$ in a cubic lattice.
Using this ansatz the single-particle Schr\"odinger equation of a particle in a
periodic potential reduces to an equation for the function $u$:
\begin{equation} \left(\frac{\hbar^2}{2m} (-i\nabla+ \k)^2
	+V_L(\r)\right)u_\k^{(n)}(\r)=E_\k^{(n)}u_\k
	^{(n)}(\r)\,,\label{eq:bloch} \end{equation} Thus, the original problem
of the Schr\"odinger equation on the entire volume can be split into
independent equations on a single unit-cell for each quasi momentum. There are
well established analytical approaches \cite{Ashcroft1956} to treat
Eq.~\eqref{eq:bloch}. A numerical diagonalization is conveniently achieved by
expanding of the periodic functions $V(\r)$ and $u_\k^{(n)}(\r)$ in discrete
Fourier sums. Keeping only the first few terms of the Fourier sums is
sufficient to obtain accurate results.

Energy structures for a one-dimensional lattice are shown in Fig. \ref{fig:bh}.
For weak periodic potentials the energy structure resembles the backfolded
parabolic single-particle spectrum. However, small gaps open at the band
crossings at the center and the boundaries of the Brillouin zone. Increasing
the potential height the gaps become more pronounced which leads to a
flattening of the energy bands. At large potential height, the spectrum
approaches the one of bound states in a harmonic potential.

It is useful to introduce Fourier transforms of the Bloch functions over the
first Brillouin zone, the Wannier function
$w^{(n)}(\r-\R)=\frac{1}{\mathcal{V}_0}\int_{\k \in 1\text{st BZ}}d^3 k
\,e^{-i\k\cdot\R}\phi_\k^{(n)}(\r)$ of a chosen lattice site $\R$.
$\mathcal{V}_0$ is the volume of the $1\text{st}$ BZ. The Wannier functions are
orthonormal and form a complete basis set. In particular, in a reasonably deep
optical lattice, the Wannier functions are localized around the lattice site
$\R$.

\subsection{Derivation of lattice models}
\label{sec:KollathLatticeModels}
In this section we will derive a lattice description of interacting ultracold
atoms in optical lattice potentials. For simplicity we focus on a one-component
Bose gas with contact interactions. However, the derivation is easily
generalized to other situations such as fermionic atoms, multicomponent mixtures, or
long-range interaction. 

The many-body Hamiltonian for ultracold bosons in a periodic optical potential
can be written as 
\begin{eqnarray} \label{eq:h_cont} \hat H=\!\!\int\!
	\textrm{d}^3r \left[ \hat \psi^\dagger(\r) \left(-\frac{\hbar^2}{2m}
		\nabla^2+V(\r)\right) \hat\psi(\r) + \frac {g} 2 \left(\hat
	\psi^\dagger(\r)\right)^2 \!\left( \hat \psi(\r) \right)^2 \right]\!.
\end{eqnarray}
The bosonic field operators $\hat \psi(\r)$ and $\hat \psi^\dagger(\r)$ obey
the bosonic commutation relations $\left[\hat \psi(\r),\hat
\psi^\dagger(\r')\right]=\delta(\r-\r')$. A pseudo potential of strength
$g=\frac{4\pi\hbar^2 a_s}{m}$ with s-wave scattering length $a_s$ is used to
describe effectively the involved interatomic potential
\cite{Pethick2002,Bloch2008}\footnote{See chapter 4 for
a detailed discussion of effective interaction potentials.}. Here $V(\r)$ is an
external potential. In the case of a periodic potential $V(\r)=V_L(\r)$, it is
appropriate to expand the bosonic field operators in terms of the Wannier
functions $\hat\psi(\r)= \sum_{\R,n} w_n(\r-\R)\hat b_{\R,n} $. The operator
$\hat b_{\R,n}$ annihilates an atom in the n-th Bloch band at a lattice site
$\R$ and fulfils $\left[\hat b_{\R,n},\hat
b_{\R',n'}^\dagger\right]=\delta_{\R,\R'}\delta_{n,n'}$. Using this expansion,
a lattice representation of the many body Hamiltonian \ref{eq:h_cont} can be
derived. In general many different terms exist with both inter- and intraband
connections. However, in the situation that all energies are much smaller than
the separation of the Bloch bands, the description can be confined to the
	lowest Bloch band $n=0$. For deep optical lattices this separation can be obtained by approximating each well by a harmonic oscillator potential. For these the band separation is roughly $\hbar\omega_\nu=2\sqrt{E_R
	V_{\nu}}$.
In the following, the index $n=0$ will be
	dropped for notational simplicity.
The Hamiltonian then reads
\begin{eqnarray}
	\label{eq:bhm_general}
	\hat H=-\!\!\sum_{\R,\R'}\!\!\!\left(J_{\R,\R'}\hat b^\dagger_\R\hat b_{\R'}\!+\text{h.c.}\right)\!+ \frac 1 2 \!\!\!\!\sum_{\bfrac{\R,\R'}{\R'',\R'''}}\!\!\! U_{\R,\R',\R'',\R'''} \hat b^\dagger_\R \hat b^\dagger_{\R'}\hat b_{\R''}\hat b_{\R'''} ,
\end{eqnarray}
with 
\begin{eqnarray}
J_{\R,\R'}&=&
-\int {\textrm d}^3r\; w^*(\r-\R)  \left(-\frac{\hbar^2}{2m} \nabla^2 +V_{L}(\r)\right)w(\r-\R')\notag\label{eq:jfull}\\
&=&- \frac {1}{\mathcal{V}_0} \int_{\k\in 1\text{st}\text{ BZ}}d^3k \, e^{i\k(\R-\R')}E_\k \quad \textrm{and} \label{eq:jek} \\ U_{\R,\R',\R'',\R'''}&=& g\!\! \int \!\!{\textrm d}^3r\; w^*(\r-\R) w^*(\r-\R') w(\r-\R'') w(\r-\R'''). 
\end{eqnarray}
The band energy $E_\k$ and the Wannier functions are obtained from the Bloch
equations for a single particle \eqref{eq:bloch}. It is important to note that
the effective hopping amplitudes require solely the evaluation of the band
energies, whose calculation by analytical or numerical means is much more
straightforward than for the Wannier functions.

For sufficiently strong optical lattice potentials, the dominating terms are
given by the nearest neighbour hopping processes with amplitude
$J_\nu=J_{\R,\R+{\bf e}_\nu}$ and the on-site interaction of strength
$U=U_{\R,\R,\R,\R} $. In addition, the non-periodic part of the potential can
be treated perturbatively and leads to an inhomogeneous slowly varying trapping
potential $V_T(\R)=-J_{\R,\R}$. Considering only these terms the Hamiltonian
reduces to the well known Bose-Hubbard Hamiltonian
\cite{Bradley1984,Fisher1989}: \begin{eqnarray} \label{eq:bhm} \hat H=\sum_{\R}
	\left[-\!\!\!\!\sum_{\nu=x,y,z}\!\!\!\left(J_\nu\hat b^\dagger_\R\hat
	b_{\R+\bf{e}_\nu}\!\!+\text{h.c.}\right)+\frac U 2 \hat n_\R(\hat
n_\R\!-\!1)+V_T(\R)\hat n_\R\right].  \end{eqnarray}

For for a deep cubic optical lattice ($V_\nu >> E_r$), an approximate
expression for the hopping amplitudes can be obtained from an asymptotic
solution of the Mathieu equations giving $J_\nu\approx
\frac{4E_R}{\sqrt{\pi}}\left(V_\nu/E_R\right)^{3/4}
e^{-\left(4V_\nu/E_R\right)^{1/2}}$. The interaction
strength $U$ can be evaluated assuming a Gaussian form of the Wannier functions
(harmonic approximation)\cite{Bloch2008} which leads to $U=g\int \mbox{d}r^3
|w_0(\r)|^4=\frac{4\sqrt{2\pi}a_s}{\lambda}\left(E_R V_xV_yV_z\right)^{1/4}$.

In summary, by using the Wannier basis the original Hamiltonian in continuum
space is reduced to a lattice Hamiltonian with only few effective parameters.
The properties of the Wannier functions come into play explicitly only when
relating lattice observables to experimental quantities in continuum space. One
example are the time-of-flight measurements which are used to access the
momentum distribution in the lattice. The absorption image at a position $\r$
after free expansion for a sufficiently long time $t$ is given by $n(\r)\propto
\left(\frac{m}{\hbar t}\right)\abs{\int d r' e^{i\k\cdot \r'}
w(\r')}^2\sum_{\R,\R'}e^{-i \k\cdot(\R-\R')} \av{\hat b^\dagger_\R \hat
	b_{\R'}}$, where $\k=\frac{m\r}{\hbar t}$ \cite{Bloch2008}. For shorter
	time of flight corrections to this relation have to be taken into
	account \cite{Pollet2012}.

\section{Superfluid to Mott-insulator transition}
\label{sec:KollathPhases}
Tight-binding models such as the previously derived Bose-Hubbard model play an
important role in condensed matter physics. This is due to the fact that they
are the simplest lattice models treating the competition of kinetic and
interaction energy. Typically this competition leads to quantum phases both of
localized and delocalized nature. In the following we are going to discuss
these phases in the context of ultracold bosons in optical lattices.

We examine interacting bosonic atoms on a simple cubic lattice. We start with a
qualitative discussion of the quantum phases occurring at zero temperature.
Further we introduce the Gutzwiller treatment, which is one of the simplest
methods available to capture the essential physics of the superfluid to
Mott-insulating transition at zero temperature. This approach is quite accurate
at large coordination numbers $z$. However, when lowering dimensionality
quantum fluctuations become more important and methods beyond mean field are
required. We focus mainly on ideal homogeneous systems. The influence of
confinement potentials is studied in Sec.~\ref{sec:trap}.

\begin{figure}[tb!]
  \begin{center}
    \includegraphics[width=1.03\textwidth]{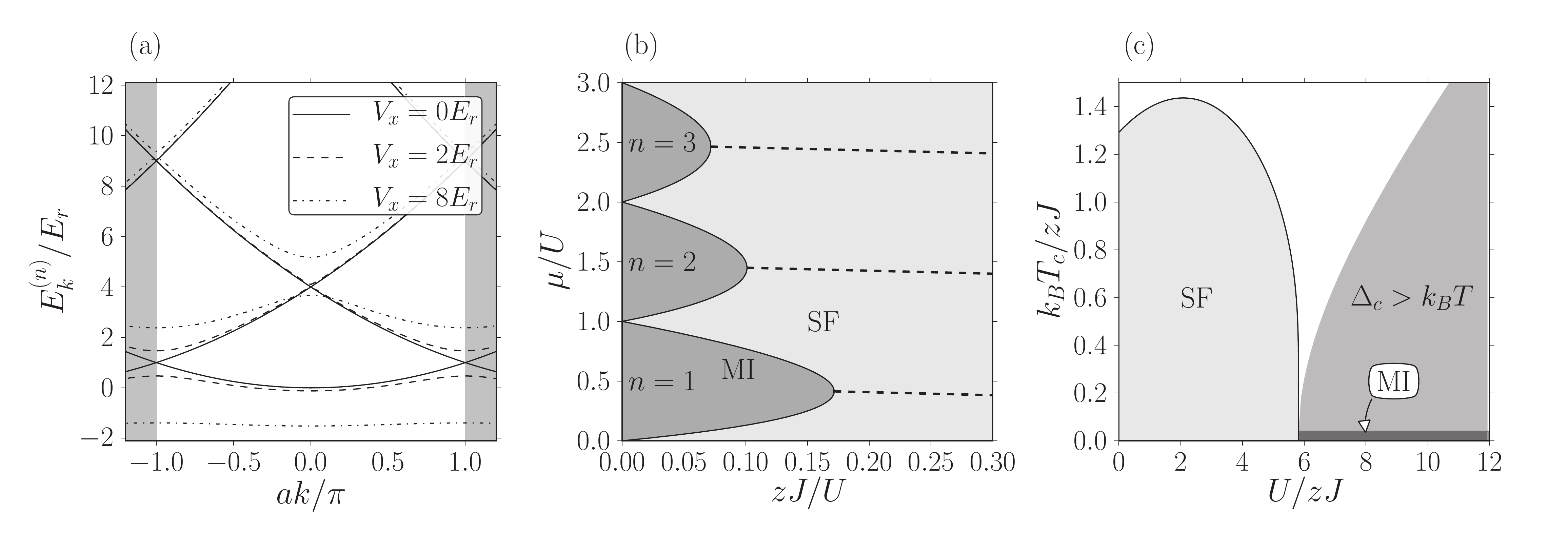}
  \end{center}
  \caption{\label{fig:bh}(a) Dispersions in a one-dimensional optical lattice
  determined from the Bloch equations \eqref{eq:bloch}. The 1D potential is
  $V(x)=V_x\sin^2(k_L x)$ and all energies are in units of the recoil energy.
  Grey regions are outside the first Brillouin zone. At $V_x=0$ the energy
  bands are connected at $k=0$ and $|k|=\pi/a$. The periodic potential leads to
  the splitting of bands at the band crossings. (b) The mean field ground state
  phase diagram of the Bose Hubbard model as a function of interaction and
  chemical potential. The dark regions mark Mott lobes at different fillings
  (MI), the bright region corresponds to the superfluid phase (SF). The dashed
  lines represent constant integer fillings which enter the Mott phase at the
  multicritical points. (c) Finite temperature phase diagram at filling one.
  The superfluid region is separated from the normal phase by a phase
  transition. The phase boundary is determined by a mean field theory
  \cite{Dickenscheid2003}. The Mott-insulator exists strictly speaking only at
  zero temperature. At finite temperature, the gap of the Mott phase is
  sketched symbolically which can be used as a line of crossover between the
  Mott phase at $T=0$ and the normal phase.}
\end{figure}
\paragraph{Limiting cases --}
It is instructive to consider the Bose-Hubbard model first in the two limits of
vanishing interaction strength $U=0$ or vanishing hopping amplitude $J=0$.  We
consider the model on a $d$-dimensional hypercubic lattice with volume
$\mathcal{V}$.  $N$ labels the total number of bosons and the average number of
particles per site is $\bar n=N/\mathcal{V}$.  In the non-interacting limit,
$U=0$, the ground state is the condensate $$\ket{\psi_0(U=0)}=\frac{\left(\hat
	b^\dagger_{\bk=0}\right)^{N}}{\sqrt{N!}}\ket{0},$$ where $\ket{0}$ is
	the vacuum. In this state the atoms are delocalized and thus local
	number fluctuations are large. The state is compressible. In two and
	three dimensions, a small interaction directly leads to the formation
	of a superfluid state with a linear dispersion at low energy.

In the opposite limit, $J=0$, the so-called atomic Mott-insulator
\begin{equation}
\label{eq:atomicMI}
\ket{\psi_0(J=0)}=\ket{\bar{n},\dots,\bar{n}}=
\prod_\R\frac{\left(b^\dagger_\R\right)^{\bar{n}}}{\sqrt{\bar n!}}\ket{0}
\end{equation}
 is
the unique ground state at integer filling $\bar n$. The atoms are localized on
each lattice site in order to minimize the interaction energy and number
fluctuations are completely suppressed. Elementary low energy excitations
consist of pairs of localized defects with occupancies of $\bar n -1$ and $\bar
n+1$ atoms. These are separated from the ground state by a gap of energy $U$
which causes the incompressible nature of the state. 

In the thermodynamic limit, the two extreme cases of the superfluid and Mott-insulator are connected by a quantum phase transition. Away of integer filling, there is no unique ground state in the limit of $J=0$ and the system remains a delocalized superfluid at any finite hopping strength and interaction.

\paragraph{Quantum phase transition -- }
One of the simplest approaches which uncovers qualitatively the phase
transition between the superfluid and the Mott-insulator is the so-called
Gutzwiller approach (see e.g.~Ref.~\cite{Jaksch1998}). It is a
variational approach and relies on the trial wave-function
$\ket{\psi_G}=\bigotimes_\R \ket{\phi_\R}$ which decouples the different
lattice sites. This ansatz considers kinetic exchange processes between
different sites only on the mean field level and neglects possible non-trivial
correlations. The resulting ground state becomes exact in both limits $U=0$ and
$J=0$. 
The wave functions $\ket{\phi_\R}$ can be expanded in the Fock basis of the single site, i.e. $\ket{\phi_{\R}}= \sum_n f^{(\R)}_n \ket{n}$ with $n=0,\dots,N$. In order to obtain the best approximation for the ground state wave function, the coefficients $ f^{(\R)}_n$ are determined by a minimization of the energy per site  $e=\frac 1 {\mathcal V} \bra{\psi_G}H \ket{\psi_G}$ while fixing the average total particle number by the introduction of a Lagrange multiplier, the chemical potential $\mu$.

In the case of a homogeneous system, the coefficients are identical for all lattice sites and we drop the label $\R$. This simplifies considerably the minimization problem of the energy
\begin{equation}
	e=-zJ\abs{\sum_n f_n^*f_{n+1}\sqrt{n+1}}^2+\sum_n\left(\frac{U}{2} n(n-1)-\mu n\right)\abs{f_n}^2\,,\label{eq:guwi}
\end{equation}
where $z=2d$ for the hypercubic lattices. An identical energy functional can be obtained by a mean-field decoupling of the kinetic term of the Hamiltonian \cite{SachdevBook}.

For a general set of parameters, one can perform the minimization of Eq.~\eqref{eq:guwi} numerically. Additionally, the location of the transition line between the Mott-insulator and the superfluid can be approximated analytically. Since the Mott phase has $f_{\bar n}=1$ and $f_{n}=0$ for $n\neq \bar n$, at the transition point one investigates the stability of  small particle and hole excitations on top of the Mott state. To be specific, we restrict to the particle-hole symmetric case for which the non-vanishing coefficients can be parametrized by a single coefficient $\alpha$: $f_{\bar{n}}=\sqrt{1-\alpha^2}$ and $f_{\bar{n}\pm 1}=\alpha/\sqrt{2}$.
The energy to leading order in $\alpha$ is given by the expression,
$$e\approx U/2 \bar{n}( \bar{n}-1)-\mu \bar{n} +\frac{\alpha^2}{2}\left(U-zJ
\abs{\sqrt{\bar{n}}+\sqrt{\bar{n}+1}}^2\right),$$  where the first two terms,
correspond to the energy in the atomic Mott-insulator. Therefore, a finite
condensate fraction ($\alpha\neq0$) becomes favorable if the term quadratic in
$\alpha^2$ is negative, thus $\Jcrit=\abs{\sqrt{\bar{n}}+\sqrt{\bar{n}+1}}^2$.
These arguments can be extended to non particle hole symmetric cases providing
the critical line $(\mu/U)_{\textrm{c}}=-\left( zJ/U+(2\bar
n-1)\pm\sqrt{1-2zJ(1+2\bar n)/U+(zJ/U)^2}\right)/2$\cite{Fisher1989}. This
analysis corresponds to the expansion of the energy in the powers of the order
parameter in the Landau theory (see e.g.~Ref.~\cite{SachdevBook}). 

The zero-temperature mean field phase diagram is sketched in Fig.~\ref{fig:bh}.  Lobes of Mott-insulating phases are present at small $J/U$. Additionally to the phase transition, the lines of integer density are indicated. To cross the phase transition there exist two possibilities: along the equal density line across the 'tip' of the lobe or the transition via a density change. The two transitions are of different universality classes. Crossing at the tip is a multicritical point which is in the universality class of the XY-model or the O(2) quantum rotor model. In contrast, the incommensurate-commensurate crossing is of mean-field type \cite{Fisher1989,SachdevBook}. The lower critical dimension of the multicritical point is $d=1$. In a one-dimensional system, a transition between a Mott-insulator to a critical phase occurs and at the tip of the lobe this transition is of Berezinsky-Kosterlitz-Thouless type \cite{Kuhner1998,Cazalilla2011}.

Even though the Gutzwiller ansatz describes well the transition between a superfluid and a Mott-insulating phase, it has several shortcomings. Due to the simplicity of the ansatz non-trivial spatial correlations cannot be incorporated in the treatment. In particular, the factorization of the wave function implies that only condensed atoms contribute to correlations $\av{\hat b^\dagger_\R\hat b_{\R'}}=\av{\hat b^\dagger_\R}\av{\hat b_{\R'}}=n_c$  for $\R\neq\R'$, and possible short range features in the correlations are completely neglected. This means that the Mott-insulating phase is trivialized to an atomic Mott-insulator (eq.~\ref{eq:atomicMI}) and many features of the many-body phases in particular close to the transition are missed. 

Nowadays, the properties of the quantum phases and the transition are well understood also beyond mean field. Due to highly accurate numerical methods (see Sec. \ref{sec:KollathMethods}) such as the density matrix renormalization group (DMRG)  and quantum Monte Carlo (QMC) the location of the phase boundaries has been determined too a high accuracy. In Table  \ref{tbl:mottphases} we summarize the values from refined treatments. In a three-dimensional cubic lattice the exact value of the critical point is relatively close to the mean field value. In contrast, in 1D, a strong overestimation of the critical interaction value is found. 

\begin{table}[h]
\tbl{Superfluid-Mott transitions of the Bose-Hubbard model}
{\begin{tabular}{@{}cccc@{}} \toprule
	dimension & $\Jcrit$ & Method\\ \colrule
	1&0.595(1)\cite{Kuhner2000,Zakrzewski2008}& Density Matrix Renormalization Group \\
2& 0.2380(1) \cite{CapogrossoSvistunov2008}& Quantum Monte Carlo\\
3& 0.2044(1) \cite{Capogrosso2007}& Quantum Monte Carlo\\
$\infty$&$(1+\sqrt{2})^{-2}=0.172$& Gutzwiller\\
\botrule
\end{tabular}}
\begin{tabnote}
	Critical values for the Mott-to-superfluid transition for the homogeneous Bose Hubbard model at filling one. The numerical results have been obtained by different methods which shall be reviewed in Section \ref{sec:KollathMethods}.

\end{tabnote}
\label{tbl:mottphases}
\end{table}

\paragraph{Finite temperature --}
The concept of a zero-temperature quantum phase diagram such as shown in Fig. \ref{fig:bh}b is useful for theoretical considerations. In practice, however, finite temperature corrections need to be taken into account. This is especially true for cold gases, where the effective temperatures are not particularly low as compared to the energy scale of the coherent particle hopping. In Fig. \ref{fig:bh}c we show the mean-field finite-temperature phase diagram of the Bose-Hubbard model \cite{Dickenscheid2003} at filling one as a function of the interaction strength. For $U/J< \left(U/J\right)_{\text{c}}$ the superfluid phase survives up to a certain critical temperature $T_c$, which becomes of the order than the hopping sufficiently far from the critical point $k_BT/J>0$. The Mott phase does strictly speaking not survive at finite temperatures. Its characteristic feature, the absence of fluctuation, or the vanishing compressibility, respectively, is limited to zero temperature. Nevertheless, it is common to speak of the Mott regime for $T<<\Delta_c$, where $\Delta_c$ is the (zero-temperature) Mott gap, since the fluctuations remain exponentially suppressed. This region is drawn symbolically in Fig \ref{fig:bh}c.

\subsection{Fermi-Hubbard model}
For fermionic quantum gases loaded into an optical lattice the reduction to
tight binding Hamiltonians is analogous to the bosonic case \cite{Bloch2008}.
Consequently, a two-component mixture with short range interactions in a
sufficiently strong lattice potential leads to the prominent Fermi-Hubbard
model 
\begin{eqnarray}  
	H \!\! &=& \!\! -J\!\!\!\sum_{\R,\nu=x,y,z}
      \!\left(\hat c_{\R ,\sigma}^\dagger \hat c^{\phantom{\dagger}}_{\R+\bf{e}_\nu,\sigma}+\mbox{h.c.}\right)
      + U\!\sum_{\R} \hat{n}_{{\R},\uparrow} \hat{n}_{{\R},\downarrow}\,, \nonumber
\end{eqnarray}
where $\hat c_{\R,\sigma}^\dagger$ is the creation operator for a fermion with
(pseudo-) spin $\sigma=\uparrow\!,\downarrow$ (typically a hyperfine degree of
freedom) and site index $\R$. The operator $\hat{n}_{\R,\sigma} = \hat
c_{\R,\sigma}^\dagger \hat c^{\phantom{\dagger}}_{\R,\sigma}$ is the density
operator. The parameters, $J$ the hopping coefficient and $U$ the interaction
strength, can be obtained analogously to the bosonic case (see Sec. \ref{sec:KollathLatticeModels}) by an expansion of the annihilation and creation operators  of each species with the help of Wannier functions. 
(See Sec. \ref{sec:KollathLatticeModels}).

The Fermi-Hubbard model is considerably richer than its bosonic analogue.
The spin degree of freedom and the fermionic statistics lead to intriguing
quantum phenomena. Examples reach from liquids over Mott-insulators and antiferromagnets to paired superfluids \cite{GebhardBook,EsslerBook,Lee2006}.
Summarizing the theoretical
investigations concerning the various possible phases of the Fermi-Hubbard
model goes beyond the scope of this review. In particular since not all
properties are fully clarified, as for example the phase diagram of the doped
two-dimensional and anisotropic three-dimensional Hubbard model.  Here, we
would like to focus on the Mott transition and the emergence of the
antiferromagnetic (N\'eel) order which are within reach in current experimental
cold atom setups. 
For a more detailed description of the
phases in the context of cold gases we refer to Ref.~\cite{GeorgesLesHouches}.

At strong repulsive interaction and half filling a gap opens in the charge
sector of the excitation spectrum of the Fermi-Hubbard model. The origin of
this charge gap is similar to the previously discussed bosonic case. It can be
easily understood at large interaction strength. Due to the strong interaction,
charge fluctuations are suppressed and a Mott-insulating state occurs in which
charge degrees of freedom are localized on single lattice sites. The lowest
charge excitations are particle-hole like and cost approximately an energy
$\Delta_c\sim U$. Thus, a crossover is induced between a liquid at low
interaction strength and a Mott-insulating state at strong interactions. At
finite temperatures much smaller than the charge gap, the characteristic
suppression of charge fluctuations of the Mott-insulating state persist.   

In addition to the charge modes, spin degrees of freedom are present. These
spin degrees lead to a highly degenerate ground state in the limit of
$U/J\rightarrow \infty$. This degeneracy is lifted at a lower interaction
strength, where an effective magnetic coupling between spins emerges. 
At large interaction $U\gg J$ this is due to the so-called superexchange process:
A second order hopping process of neighbouring fermions
via an intermediate highly energetic doubly occupied state. Due to the Pauli
principle the doubly occupied state is only possible for fermions of
different spin. Consequently, the arising effective coupling is
antiferromagnetic and the coupling strength
is given by $J_{ex}=\frac{4J^2}{U}$ \cite{GeorgesLesHouches}.

\begin{figure}[tb!]
  \begin{center}
    \includegraphics[width=0.44\textwidth]{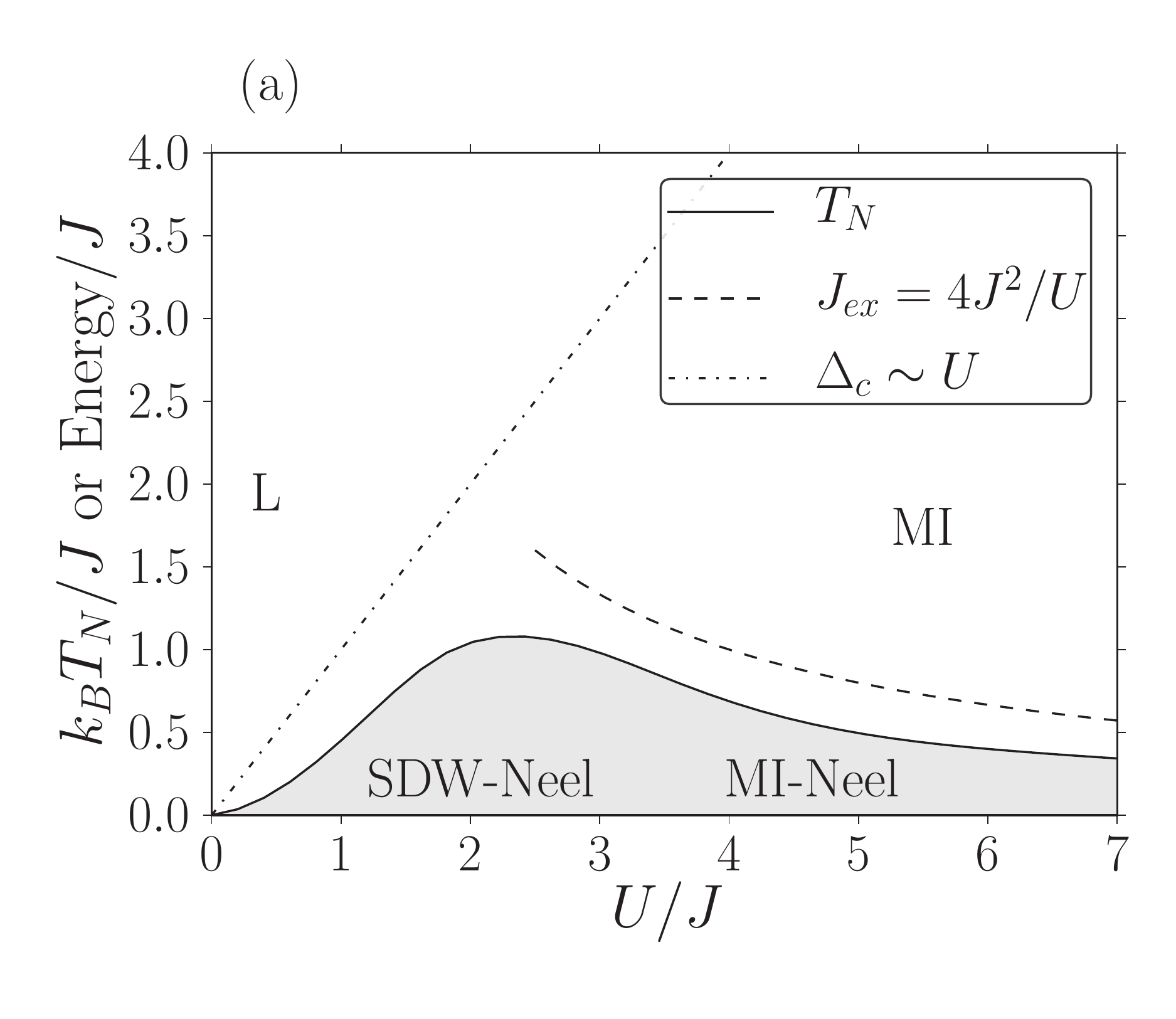}
    \includegraphics[width=0.55\textwidth]{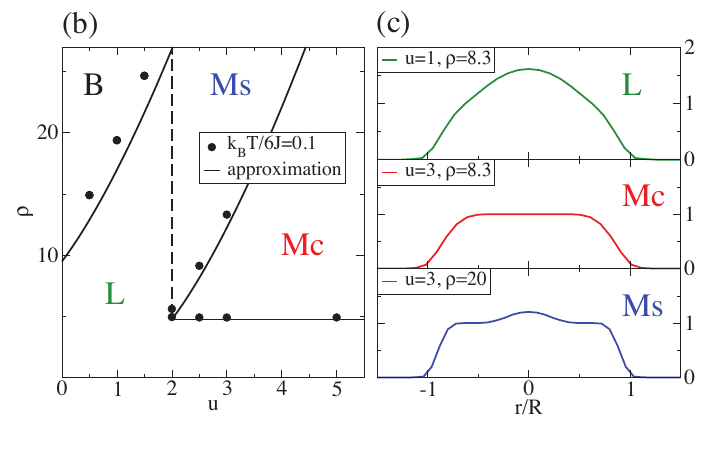}
  \end{center}
  \caption{\label{fig:fh} 
(a) Qualitative phase diagram of the Fermi-Hubbard model in three dimensions at
half filling as a function of interaction $U$ and temperature \cite{GeorgesLesHouches}. The
N\'eel-ordered antiferromagnet at low temperatures is understood as a
spin-density wave (SDW-N\'eel) at weak interactions and a Mott-antifferromagnet
(MI-N\'eel) at strong interactions. The latter is characterized by the
superexchange constant $J_{ex}$. The charge gap $\Delta_c$ can be used to
distinguish between the liquid (L) and Mott-insulating (MI) regimes in the
normal phase at temperatures beyond the N\'eel temperature $T_N$. (b) State
diagram using LDA with a parabolic trapping potential as a
function of the characteristic density $\rho$ and effective interaction $u=U/J$. Symbols were
obtained using DMFT and lines are an analytical continuum approximation for the case $T=0$ (from Ref. \cite{DeLeo2008}). There are four regimes: B (band insulator in the center of
the trap), Mc (Mott insulator in the center of the trap, shaded areas), Ms
(shell of Mott insulator away from the center), and L (liquid state). The solid lines
indicate the crossovers between the regimes. The vertical dashed line represents the crossover from the liquid to
the Mott state. Examples for density profiles plotted with respect to the relative radius $r/R$ are shown in (c). These profiles are taken at values of
$\rho$ for which the central densities are 0.995, 1.005, and 1.995 (from bottom
to top). (b) and (c) are from Ref.~\cite{DeLeo2008}.}
\end{figure}

In three dimensional cubic lattices, the superexchange coupling induces a phase
transition to an antiferromagnet with long-range order at low
temperature (Fig.~\ref{fig:fh} (a)). The dome-like structure of the
phase-boundary can be understood intuitively: At large
interaction the energy scale for the antiferromagnetic coupling reduces as
$J^2/U$ leading to a decreasing transition temperature with increasing
interactions. In contrast, at low interaction, the charge gap 
becomes small and charge fluctuations can destroy the magnetic ordering. In
this regime of weak repulsive interaction, antiferromagnetic order is due to a
spin-density-wave transition, in which the opening of the insulating gap and
antiferromagnetic order occur simultaneously.

Due to the Mermin-Wagner theorem, N\'eel order at finite temperature is
restricted to the 3D case. In two-dimensional lattices long-range antiferromagnetic order exists only
in the limit $T=0$. In 1D, the ground state at half-filling in the presence of
repulsive interaction is a Mott insulator with algebraically decaying
antiferromagnetic ordering. Ground state and finite temperature properties of the 1D 
case can be obtained
analytically from the Bethe-ansatz solution \cite{EsslerBook}.

The first experiments with non-interacting fermionic atoms loaded into a cubic
optical lattice observed the change of the Fermi
surface with increasing the number of
atoms\cite{KoehlEsslinger2005,Esslinger2010}. Later, the characteristic
suppression of particle fluctuations in the Mott-insulating phase has been
detected \cite{Schneider2008,JoerdensEsslinger2008}. Only recently the first
signs of short range antiferromagnetic correlations have been discovered by
modulation spectroscopy \cite{Kollath2006,GreifEsslinger2011} and by using a superlattice
\cite{Greif2012}.
However, the spontaneous formation of long range antiferromagnetic order has
not been realized so far. This is due to the relatively high temperatures, of
the order of the hopping amplitude, which are currently present in these
fermionic gases experiments \cite{Jordens2010}. One of the main challenges is
therefore the design of efficient cooling schemes \cite{McKay2011} to reach the
interesting low temperature phases. Even lower
temperatures than for the antiferromagnet would be required to address the long
standing question of unconventional superconductivity in the doped two-dimensional Hubbard
model.

\subsection{The influence of the trap}
\label{sec:trap}
Even though a lot of progress has been recently in the creation of flat bottom traps to confine the atoms, in most current experimental setups, the presence of a trapping potential has to be taken into account. The trapping potential can either result from the presence of a dipole or magnetic trap or the amplitude focus of the lattice beams itself. Since the spatial extension of the atomic cloud is small as compared to the characteristic variation of the trapping potentials, a parabolic potential represents often a good approximation. Here we will assume the more general form $V(\r)= V_t(\r/a)^\alpha$, where $\alpha$ is the exponent and $a$ the lattice constant. In order to discuss the consequences of the presence of such a potential on the observed physics, it is useful to consider the so-called local density approximation. This approximation treats the external potential as a spatially varying chemical potential, i.e. $\mu(\R)=\mu_0-V(\R)$. Here $\mu_0$ is the chemical potential in the center of the trap. Local observables in the trap $O(\R)$ are connected to their grand-canonical homogeneous counterpart $O_h(\mu)$ by the relation $O(\R)= O_h(\mu(\R))$.
Therefore, moving through the trapping potential in space corresponds to following a vertical line in a grand-canonical phase diagram (e.g.~Fig. \ref{fig:bh}b for bosonic atoms). Depending on the interaction strength, this can lead to the coexistence of conducting and insulating phases. The resulting states can be characterized by their density profiles. Within the local density approximation, the state diagram depends only on the characteristic density $\rho=N(V_t/zJ)^d/\alpha$ (and not on the total number of particles $N$ and potential $V_t$ separately) \cite{RigolScalettar2003,DeLeo2011}. An example of such a characterization is summarized in the state diagram for the fermionic gas shown in Fig.~\ref{fig:fh} (b) (see also chapter 6). This is due to the fact that all global quantities and local quantities at the center of the trap only depend on $\rho$. 
 In order to show this dependence, consider a mean value $\bar O$ which corresponds to the sum of the corresponding local observable $O(\R)$:
 $\bar O=\frac 1 N\sum_\R O_{\R}$. Within the local density approximation the observable can be related to its homogeneous counterpart $O_h$ by $\bar O=\frac{1}{N}\sum_\R O_h (\mu(\R))$. Using the continuum limit this can be expressed as 
 \begin{equation}
\label{eq:lda}
\bar O=\frac{\Omega_{d-1}}{\rho\alpha} \int_{-\infty}^{\tilde\mu_0} d \tilde\mu \, (\tilde\mu_0 - \tilde\mu)^{\frac{d}{\alpha}-1} O_h(\tilde\mu)
\end{equation}
 with  $\tilde\mu\equiv\mu/zJ$ and $\Omega_{d-1}$ being the surface of a sphere
 in $d$ dimension. From this expression we see that all such observables only
 depend on the characteristic density and the chemical potential $\tilde\mu_0$
 in the center of the trapping potential.  The chemical potential $\tilde\mu_0$  itself only depends on
 $\rho$, since $\mu_0$ can be found by using Eq. \eqref{eq:lda} for the
 particle number ($O(\R)=\av{\hat n(\R)}$). Thus all local quantities on the
 central site and the global quantities depend on $\rho$.

Deviations from the local density approximation can occur, in particular, close to the transition regions between different states. Further, the influence on more complex physical quantities such as spatial correlations has for example been detailed in Ref.~\cite{KollathZwerger2004}.

\subsection{Non-equilibrium aspects}
By the tunability of the system parameters, such as the geometry or the interaction strength, a wide variety of non-equilibrium situations can be realized with ultracold atoms in optical lattices.
These can be triggered for example by instantaneous or slow changes of global parameters, local excitations, or periodic external driving. 
The physical phenomena realized by these processes are manifold and we would like to briefly highlight a few selected ones.

The laser intensities can be changed much faster than any other time-scale of the system 
 which enables experiments in the extreme non-adiabatic limit by performing a so-called {\it instantaneous quantum quench}. Due to the minimal coupling to their environment,  a highly excited state created by a quantum quench relaxes mainly via intrinsic scattering mechanism. An early experimental manifestation of isolated system dynamics was the observation of the collapse and revival of the superfluid phase after the sudden increase of the optical lattice height \cite{Greiner2002}. 
 This first experiment in optical lattices has been followed by an explosion of interest \cite{Bloch2008}. It is very difficult to treat quantum dynamics analytically or numerically and ultracold atoms have the potential to probe time scales which go beyond the possibilities of current state-of-the-art theoretical methods \cite{Trotzky2011}. Questions under investigation concern the spreading of correlations, the existance of so-called Lieb-Robinson bounds for this spreading (See e.g. \cite{Calabrese2006, Cheneau2012} and references therein) and the relaxation dynamics with a focus on the possible thermalization \cite{Polkovnikov2011b}.

 It has for instance been confirmed experimentally, that the intrinsic relaxation mechanism in quantum systems depends strongly on the dimensionality of the system \cite{Kinoshita2006}. Also crossovers between integrable and non-integrable regimes \cite{Kollath2007,ManmanaMuramatsu2007,Polkovnikov2011b} would be interesting to be studied with ultracold atoms.

Nonequilibrium dynamics are also useful to demonstrate the presence of certain interaction effects, such as the spin exchange with two-component bosons \cite{Trotzky2008}. Another approach to study dynamical properties of a system is by exciting {\it local impurities} and to monitor locally the subsequent time evolution. This has been realized in one-dimensional tubes with gases of different bosonic species \cite{Palzer2009,Catani2012} or in an effective Heisenberg model \cite{Fukuhara2013}.

Due to the lack of a thermal bath, the understanding of non-abiabatic effects during a {\it slow quench} of a system parameter is highly important. Such a slow variation is for example used to prepare correlated quantum states from condensates. While crossing a quantum phase transition a scaling behaviour is expected for the defect creation rates. Such a universal scaling behaviour is well known from classical systems and reasoned in these by a simple picture called the Kibble-Zurek mechanism \cite{Polkovnikov2011b}. First experiments in cold atomic gases in optical lattices have addressed specifically the situation a of slow parameter change across a quantum phase transition \cite{Bakr2010,Sherson2010}. However, in experiments with trapped gases, additionally to the dynamics expected in the homogeneous counterpart, a mass redistribution takes place. In the case of strongly interacting gases the formation of insulating regions can strongly suppress the mass transport \cite{Bernier2010,NatuMueller2011,Bernier2011a} and lead to a strongly excited state as e.g. many defects in Mott plateaus \cite{Bakr2010,Sherson2010}.

{\it Periodic driving} of external parameters as e.g.~the optical lattice height or position have been used to perform spectroscopic measurements of many body dynamic correlations \cite{Bloch2008,Jordens2010}. Hereby typically the linear response of the system to the periodic driving close to the many-body resonances is analysed.  However, by a periodic driving far off these resonances, special many-body states can be stabilized. The stabilization of effective states works similar to the principles of the light shift far-off resonance which forms a conservative potential for the atoms. An example is the fast periodic tilting of the optical lattice potential which induces the localization of atoms or even the formation of a condensate at the edge of the Brioullin zone  \cite{Morsch2001} and can be employed in order to create magnetically frustrated structures \cite{Struck2011a}.

\section{Theoretical tools}
\label{sec:KollathMethods}
Detailed knowledge of microscopic parameters in ultracold atomic gases enable
stringent juxtapositions of model calculations and experiments for highly
non-trivial quantum phases. This is a quite unique situation.   In correlated
solid state materials, for example, a derivation of an exact model Hamiltonian
from first principles is in most cases very difficult. This is due to the
complex structures of the materials and the presence of defects and lattice
phonons.

On the one hand by this direct comparison of theory and experiment, the
theoretical description enables the understanding of the principles behind
occurring physical phenomena. Further, the theoretical simulations can be used
to extract remaining unknown variables from the experimental setups. In
particular temperature which is very hard to measure experimentally in an
optical lattice has been extracted for both bosons and fermions
\cite{Pollet2012,Jordens2010}. On the other hand, experiments can validate
theoretical approaches or reveal their systematic errors. Thus experiments
directly contribute to improvement and development of many-body methods. A
striking demonstration of the potential of such cross validation has been
accomplished with Monte Carlo simulations for the equation of state of a Fermi
gas at unitarity \cite{VanHoucke2012,Bloch2008}.

Many different approaches have proven to be very useful to understand the
occurring phenomena in optical lattices. Examples of approximate methods are
the high-temperature series expansion for fermions \cite{OitmaaBook}, strong
coupling expansions \cite{Freericks1994}, effective low-energy theories
\cite{SachdevBook} or mean-field descriptions in terms of slave particles
\cite{GeorgesLesHouches}.  Here we give an overview of quantitative numerical
tools suited for the treatment of correlated lattice models. Our selection
covers approaches to a large variety of situations: exact diagonalization and
matrix product state based methods which are appropriate tools for
low-dimensional systems, and the dynamical mean field theory which is most accurate
at high coordination numbers. We also review recent developments in quantum
Monte Carlo methods. The quantum Monte Carlo methods are not restricted to
specific geometries, but the sign problem prevents simulation of certain
fermionic and frustrated systems. Since non-equilibrium effects play an
important role in the context of cold atoms, we will also analyse the
applicability of the methods to treat time-dependent problems.

\subsection{Exact diagonalization}
\label{sec:ed}
For sufficiently small lattice systems, exact diagonalization methods
\cite{Weisse2008,Lauchli2011} can be applied to calculate eigenstates of the
Hamiltonian by directly solving the stationary Schr\"odinger equation $\hat H
\ket{\psi_n}=E_n\ket{\psi_n}$. In order to perform a numerical diagonalization
wave functions are encoded in a vector with real or complex elements and the
Hamiltonian can be represented by a matrix. With a suitable diagonalization
scheme, the eigenenergies and arbitrary properties of the states can be
calculated numerically exactly.

Exact diagonalizations are mainly limited by the size of the Hilbert space
which grows exponentially with the number of lattice sites $N$, e.g.~for
hard-core bosons $\mbox{dim}{\mathcal{H}}=2^N$. Incorporating lattice
symmetries and particle number conservation reduces the Hilbert space dimension
considerably \cite{Weisse2008,Lauchli2011}, but does not solve the fundamental
problem of exponentially growing complexity with system size. 

So-called Krylov subspace methods are most efficient to target the low-lying
eigenstates. These methods are based on an iterative application of the
Hamiltonian to the wave function. The most basic approach is the power method,
which generates the series $$\ket{\psi^{(K)}}=\left(\hat H-\lambda\right)^K
\ket{\psi_{\text{rand.}}}/\sqrt{||\left(\hat H-\lambda\right)^K
	\ket{\psi_{\text{rand.}}}||},$$ where $\ket{\psi_{\text{rand.}}}$ is a
	randomly drawn initial state and $\lambda$ is an upper bound to the
	highest eigenvalue of the Hamiltonian. In the presence of a gap the
	multiple application of the operator projects the initially random
	state onto the ground state, if an overlap between the initial and the
	ground state exists. This can be discovered decomposing the initial
	state into the eigenbasis of the Hamiltonian.  Somewhat more
	sophisticated Krylov schemes such as Lanczos, Arnoldi and the related
	Davidson algorithm \cite{Weisse2008,Lauchli2011}, converge particularly
	quickly. A few hundred multiplications can be sufficient to produce a
	ground state with high accuracy. 

The advantage of Krylov methods is that they only require a Hamiltonian-state
multiplication, which for a general matrix requires
$\propto\mbox{dim}{\mathcal{H}^2}$ operations or even using the sparseness of
typical lattice Hamiltonians, can be performed with
$\propto\mbox{dim}{\mathcal{H}}$ operations.  In order to address finite
temperature properties the full spectrum of the Hamiltonian has to be accessed.
This requires explicit storage of all $\left(\mbox{dim}{\mathcal{H}}\right)^2$
matrix elements, what reduces the accessible system sizes by roughly a factor
$1/2$ as compared to ground-state simulations. Standard routines perform the
diagonalization in $\propto\left(\mbox{dim}{\mathcal{H}}\right)^3$ operations.
Both, sparse and full diagonalization schemes, can be used to treat
time-dependent problems with only little computational overhead
\cite{Noack2005}.

In practice, state of the art implementations can calculate ground states of
hardcore bosons on almost 50 lattice sites or even more if only few particles
are present. More difficult to treat are systems requiring a large local basis,
such as weakly interacting bosons.  Convergence properties of exact
diagonalizations depend little on the physical properties of the system.
Therefore, exact diagonalizations are particularly important for frustrated
magnetic systems \cite{Lauchli2011} and fermionic problems, which often cannot
be treated efficiently by Quantum Monte Carlo methods. However, the unavoidable
finite size effects play a role in particular when quantum critical phenomena
or competing phases shall be investigated at low temperature. 
 
\subsection{Matrix product states and the density matrix renormalization group methods}
\label{sec:mps} In order to overcome the problem of the exponential growth of
the Hilbert space with the size of the system, different approaches have been
designed. For low-dimensional systems, variational methods based on so-called
matrix product states (MPS) such as the density matrix renormalization group
(DMRG)\cite{White1992} have been particularly successful. 

A general matrix product state (MPS) reads 
\begin{equation}
\ket{\psi}=\sum_{n_1,\dots,n_N}\tr A^{n_1}A^{n_2}\cdots
A^{n_N}\ket{n_1,n_2,\dots,n_N}\,,\label{eq:mps} 
\end{equation}
where $n_j=0,\dots,d-1$
label the local basis states, and for each basis element $n_j$ a matrix
$A^{n_j}$ of $D\times D$ dimensions has been introduced. Any wave function can
be represented in this form if the dimension of the matrices is taken large
enough. The efficiency of MPS is due to the fact that for 1D chains, and to
some extent also 2D lattices, the physical state can be approximated to a very
good extend using reasonably treatable values of $D$ which scale moderately
with the system size. 

The MPS construction and the requirements on the bond dimension $D$ can be
understood from the Schmidt decomposition for a partition of the lattice in
left and right subsystems. It reads $\ket{\psi}=\sum_{\gamma=1}^{D}
\lambda_\gamma \ket{\psi_L}_\gamma \ket{\psi_R}_\gamma$, where
$\ket{\psi_{R,L}}_\gamma$ are wave functions of the left and right part and
$\lambda_\gamma$ the Schmidt values connecting them. In principle, for an exact
representation of the wave function, $D$ should scale as the dimension of the
Hilbert space. However, if the Schmidt values are rapidly decaying, a limited
number $D$ is sufficient to reproduce states quasi exactly. For example in a
one-dimensional system at equilibrium and away from criticality, the required
dimension $D$ in order to represent the physical state is found to be
independent of the size of the system. This can be understood employing the von
Neumann entropy $S=-\sum_\gamma \lambda_\gamma^2 \log_2 \lambda_\gamma^2$
between the left and right subsystem as a suitable measure for the decay of the
Schmidt values. Since the area law states that the entanglement between left
and right subsystems is determined by the surface between them, the required
number of Schmidt values in order to represent well the physical state does not
increase with system length.  Iteratively applying the Schmidt decomposition
for different partitions of the system, directly leads to the desired MPS
\eqref{eq:mps} and the constituent matrices are of dimension $D$
\cite{Schollwock2011}. Consequently a Matrix product representation is highly
advantageous in comparison with the exponential demand of resources of exact
diagonalization.

The MPS scheme has its origin in the DMRG algorithm, proposed by White in 1992
\cite{White1992}, which effectively implements an iterative Krylov scheme in
the variational space of MPS. The standard DMRG algorithm is very efficient.
Ground states for chains with thousands of sites and fairly large cylindrical
system, a two-dimensional system with periodic boundary conditions along one direction and open along the other, can be calculated using DMRG, e.g. bosons on $10\times100$ sites are
realistic. Inhomogenities in systems as trapping potentials or disorder
potentials can be taken into account quasi-exactly, without resorting to the
local density approximation. Carefully extrapolating with respect to $D$ and
the system size, characteristic critical exponents can be extracted. In order
to assess the convergence of the DMRG algorithm, one has to be aware of the
fact that the method is variational and states can be trapped in local minima.
Beyond DMRG, there exist numerous variants of matrix product methods, including
algorithms for finite temperature or for dynamical properties (See e.g. Ref.
\cite{Schollwock2011}).

Also non-equilibrium situations can be addressed using MPS
\cite{Daley2004,Feiguin2005,Schollwock2011}.  In this case, however, the area
law does not apply and computational complexity can grow exponentially in time.
Therefore, the method can only be considered as accurate up to a certain time.
Dynamics of MPS with a finite $D$ deviates approximately exponentially with
time \cite{GobertSchuetz2004} and therefore extrapolations with respect to $D$
are prohibited. 

\subsection{Quantum Monte Carlo methods}
\label{sec:qmc}
Monte Carlo techniques are probably the most successful approach to simulate
generic problems in statistical mechanics \cite{LandauBinderBook}.  The main
idea is to approximate observables via statistical averages over configurations
which are sampled according to their weight in the partition sum.  The most
common scheme for realizing this importance sampling is the Metropolis
algorithm.  In its simplest form the algorithm starts from a random initial
configuration. Then a sequence of local updates is performed which are accepted
by probabilities given by relative Boltzmann weights of the configuration
before and after the update. By this different trajectories of configurations
are obtained.  One of the big challenges in the design of an efficient Monte
Carlo scheme is to find an efficient ergodic sampling of the configurations.
This is especially important close to second order phase transitions, where
diverging autocorrelation times occur, leading to so-called critical slowing
down. Non-local cluster updates have been developed to overcome this problem.

Monte Carlo methods for quantum systems (QMC) are based on mappings of the
quantum system to a classical one, for example by going over to a path integral
representation. A systematic problem occurring in certain frustrated or
fermionic problems is that the effective statistical weight of a configuration
can become negative. This is the infamous sign problem which represents the
biggest challenge in QMC developments.

Particularly successful QMC schemes in the context of cold atomic systems are
based on the expansion of path integrals in a coupling parameter which operate
on a continuous time axis. For the implementation of the Markov chain, the
jumps between different orders require special treatment. To this purpose,
so-called worm updates have been developed \cite{Pollet2012}. These schemes do
not suffer from critical slowing down and the computational effort scales
linearly with system size and inverse temperature. The worm algorithms can be
applied to inhomogeneous situations and are therefore the ideal tool for
ab-initio calculations for trapped cold atoms. Bosons in 3D optical lattices
with more than $10^5$ particles can be simulated using the worm algorithm
\cite{Trotzky2010a}. Multi-species bosons or spins are equally accessible.
Fermions at half-filling can be simulated down to the N\'eel temperature by
using determinantal Monte Carlo \cite{Pollet2012}. Away from half-filling, but
also for frustrated spin systems, the sign
problem prevails.  Also non-equilibrium dynamics exhibit a sign problem and are
difficult to treat by quantum Monte Carlo methods.

\subsection{Dynamical mean field theory}
\label{sec:dmft}
A successful route to treat strongly correlated fermions is the dynamical mean
field theory (DMFT) \cite{GeorgesRozenberg1996}. This method considers
interactions on a single site (or a small subsystems) exactly and approximates
the remaining system as an effective medium with a self-consistently determined
spectral function.  In contrast to conventional static mean field methods, DMFT
allows the dynamic exchange of particles in the small subsystem with the
reservoir. The approximation consists in assuming a local self-energy.  It can
be shown that the DMFT approximation becomes exact in the limit of infinite
dimensionality. Noninteracting and strongly interacting limits are exactly
reproduced for arbitrary lattices. Increasing cluster sizes, the exact result
can be approached systematically also for lower dimensions. 

The DMFT self-energy has to be determined from an impurity problem for the
single site or the cluster \cite{GeorgesRozenberg1996}.  The solution of this
impurity problem is computationally difficult and different approaches can be
adopted. Exact diagonalization, DMRG, and numerical renormalization group
methods can be applied, which have typically a limited frequency resolution.
Continuous time Monte Carlo schemes are more efficient, but can access spectral
functions only via analytical continuation. 

DMFT has been employed successfully to treat fermionic gases, for example in
order to determine their temperature \cite{Jordens2010}. Inhomogeneities can be
taken into account via the local density approximation, but also refined
methods exist, e.g. for the parabolic trapping of cold atoms
\cite{Helmes2008,GorelikBlumer2010} or the description of oscillating order parameter \cite{Kim2011}. A particularly interesting ongoing
development are implementations of the principles of DMFT on the Keldysh
contour to address non-equilibrium problems \cite{WernerRMP}. 

\section{Perspectives}
\label{sec:KollathOutlook}
In this short review we mainly focused on basic quantum phases of quantum
gases in optical lattice potentials. We discussed their theoretical description
using tight-binding lattice models in cubic lattices and the arising quantum
phases restricting ourselves to the lowest Bloch band.  However, we would like to point out that this is just a very small part of the fascinating physics possible with quantum gases in optical lattices. To conclude, we highlight a small subjective choice of recent achievements and ongoing developments.

The first area we would like to mention is the use of more involved lattice
geometries including artificial gauge fields \cite{Dalibard2011}. The experimental setups which can be used to create special optical
lattices structures will be discussed in detail in chapter
5 of this volume. One part of the efforts is directed
towards realization of 
non-trivial band structures which can give rise to intriguing properties such as
fractional excitations and topologically protected edge states.
For example, a tunable Honeycomb lattice was engineered by adding a detuned 
collinear beam to two-dimensional square lattice \cite{Tarruell2012}. 
This lattice geometry leads to the formation of Dirac cones in the band structure.
In one-dimensional systems topologically non-trivial band structures can be 
created using superlattice potentials. In such a system
Klein tunneling across a linear band crossing of
excited bands has been observed \cite{Salger2011}. 
In another experiment a condensate moving through the Brioullin zone has been
used to measure the Zak phase of two differently
dimerized superlattices via Ramsey fringes \cite{Atala2013}. 
Another aspect of complex lattice structures is the possibility of 
magnetic frustration. Magnetic frustration can cause interesting phases as spin liquid. However, in order to realize such a spin liquid in cold atoms,
the temperature has to be further lowered compared to the current
state of the art. 
A fascinating but very ambitious goal are topologically
ordered ground states on honeycomb lattices proposed by Kitaev, which require
strongly anisotropic exchange couplings \cite{LewensteinBook}. Yet another
class of special optical lattices aim at the generation of artificial gauge
fields \cite{Dalibard2011}.

Another interesting direction are effects beyond the lowest Bloch band of the
optical lattice. Fortunately, in some cases higher-band effects do not necessarily lead to the 
breakdown of the convenient single-band picture but only modify effective parameters.
For example, large interaction for bosonic atoms in deep optical
lattices can be incorporated into the single-band picture 
via number dependent interaction potentials
\cite{Buchler2010,Mark2011}. 
More drastic changes occur in shallow optical lattice potentials,
where ferromagnetism can emerge \cite{Mathy2009,Pilati2013}.
By explicitly populating higher bands, orbital degrees of freedom can be investigated \cite{Isacsson2005,Muller2007}.
So far, metastable p- and d-wave superfluid phases have been realized in bosonic systems \cite{Wirth2010}. On the long term, the interplay of spin and orbit may be studied with fermionic atoms in higher bands of optical lattices.

\bibliographystyle{ws-rv-van}
\bibliography{references}
\end{document}